\documentclass[journal]{IEEEtran}

\ifCLASSOPTIONcompsoc
\else
  \usepackage{cite}
\fi
\ifCLASSINFOpdf
  \usepackage[pdftex]{graphicx}
\else
  \usepackage[dvips]{graphicx}
\fi
\usepackage[cmex10]{amsmath}
\begin{document}
%
\title{Discovering and Predicting Temporal Patterns of WiFi-interactive Social Populations $^{\ast}$}

\author{Xiang~Li,~\IEEEmembership{Senior Member,~IEEE}, Yi-Qing~Zhang,~\IEEEmembership{Student Member,~IEEE}, Athanasios~V.~Vasilakos,~\IEEEmembership{Senior Member,~IEEE} 

\IEEEcompsocitemizethanks{\IEEEcompsocthanksitem Xiang Li and Yi-Qing Zhang are with Adaptive Networks and Control Lab., Department
of Electronic Engineering, Fudan University, Shanghai 200433, China, (E-mail:\{lix,,12110720032\}@fudan.edu.cn). Athanasios V. Vasilakos is with Department of Electrical and Computer Engineering, National Technical University of Athens, Athens, Greece, (E-mail:th.vasilakos@gmail.com).\protect\\

$\ast$ All correspondence should be addressed to Xiang Li.\protect\\

The paper is the invited chapter of book ``Opportunistic Mobile Social Networks'' Jie Wu and Yunsheng Wang (Eds.) published by CRC Press in Aug. 2014.
}
\thanks{}}



\maketitle

\IEEEdisplaynotcompsoctitleabstractindextext
\IEEEpeerreviewmaketitle

\section{Introduction}

Since the publication of Moreno and Jennings' sociometry book in 1934 \cite{moreno1934shall}, network (graph) theory has become one of the most powerful tools to characterize social interactions and population dynamics \cite{Freeman1992rms,Scott2000sna,wasserman1994social}. The discoveries of small-world \cite{watts1998collective} and scale-free networks \cite{barabasi1999emergence} in the late of 1990s arouse the world-wide attention on complex networks and network science, and we have witnessed ten more years the fruitful and exciting advances to understand the hidden patterns behind complex connectivity features and characteristics of diverse large-scale networking systems. Such natural and/or man-made examples range from the Internet, the World Wide Web, biological brain networks, protein-to-protein interaction networks, power-grids, wireless communication networks to categories of social, economic and financial networks at different levels of human society. Extensive efforts and elegant attempts have been devoted to answering a fundamental question: how does the fascinating complex topological features affect or determine the collective behaviors and performance of the corresponding complex networked system \cite{Albert2002smc,barrat2008dpc,Chen2012icn,Costa2011amr,Costa2007ccn,Dorogovtsev2002en,easley2010ncm,Gross2009ant,jackson2010sen,Newman2003sfc,newman2009ni,Pastor2007esi,Wang2003cns,Wang2006cnt,wang2012nsi}. While this widely concerned key question still remains open in the fast developing field of network science, especially in the newly focused situation that the information of temporal dimension as an explicit element defines the edges of such a so-called temporal network \cite{holme2012temporal} when they are active.

Nowadays, a great deal of incredible products of flourished information and communication technologies (ICTs) are unobtrusively embedded into the physical world of human daily activities: we communicate with our friends/colleagues with emails and/or mobile phones, we make web-shopping and check out by credit cards, and travel/commute in public transportation networks with the payment of transit cards, and we surf online 24-hour everyday covered by WiFi or 3G signals everywhere. Such creative digital devices/instruments not only reshape our daily life, but also record tremendous digital traces produced by human activities. These digital records offer unparalleled opportunities to explosively digitize physical human interactions and provide fresh temporal clues to throw sight into the human behavioral patterns.

In the literature, the successful stories mainly include the assistance from the Bluetooth, the active Radio Frequency Identification (RFID), wireless sensors and the WiFi technology \cite{eagle2006reality,eagle2009inferring,cattuto2010dynamics,isella2011wic,stehle2011high,salathe2010high,cai2009crossing,chaintreau2007impact,hui2008human,karagiannis2010power,zhang2012towards,zhang2012characterizing,zhang2013temporal,zhang2013exploring}. For instance, Eagle and his colleagues  took the advantage of Bluetooth embedded in mobile phones, and collected the proxy data of person-to-person interactions in the program of Reality Mining \cite{eagle2006reality,eagle2009inferring}. Barrat et al. built a flexible framework based on the RFID technology, and recorded the volunteers' face-to-face interactions in different rendezvouses such as conference, museum, and primary school \cite{cattuto2010dynamics,isella2011wic,stehle2011high}. Salath\'{e} et al. utilized wireless sensors to trace person-to-person encounters among the members of a high school, and further evaluated the respiratory disease transmission risk in school campus \cite{salathe2010high}. Besides, many researchers explored the inter-contact intervals of mobile users' interactions by the Bluetooth and WiFi technology, and designed more efficient algorithms to improve the performance of data dissemination among a large-scale population of mobile users \cite{cai2009crossing,chaintreau2007impact,hui2008human,karagiannis2010power}. Moreover, such fruitful researches push the desire to depict human social networking populations with interdependent collective dynamics as well as the behavioral patterns.

WiFi, as one of the most ubiquitous wireless accessing techniques, has been widely deployed in human daily circumstances. Actually, the
``WiFi-Free'' sign can be found in nearly every corner of urban areas, and the notion of ``WiFi-City'' becomes reality. For example, since 2010, it has been reported the significant increase of global number of WiFi certified product launches and hotspots (both public and private), while more than $9,000$ product certifications have been issued by the WiFi Alliance, and the global number of hotspots is over 280 millions \cite{wba2011}. Therefore, the flood of commercial WiFi systems comes as a powerful proxy tool to collect digital access traces of a huge population equipped with of WiFi devices. As a snapshot of the modern society, a university is in the coverage of WiFi signals, where the WiFi system records the digital access logs of the authorized WiFi users when they access to the campus wireless service. Such WiFi access records, as the indirect proxy data of a large-scale population's social interactions without artificial interference, are the targets to explore in this chapter.

As the response of the above motivations, this chapter comes as a synthetical review of our extensive efforts devoted to temporal networks and human population dynamics in the past years \cite{zhang2012towards,zhang2012characterizing,zhang2013temporal,zhang2013exploring,Cui2013cct,Pan2013sctn}, and we target a cyber-social population with the example of a university campus having the WiFi coverage. We conduct the study based on a dataset from a Chinese University, whose findings have also been verified with other open-source data sets in different rendezvous, and this chapter only reviews this series of works in detail with the focus of WiFi interactions. The left part of this chapter is organized as follows. As a preliminary section, Section \ref{background} introduces the background of the WiFi data-collection setting in an involved Chinese university, \emph{i.e.}, Fudan University,  and defines the related concepts as well as the dataset preprocessing. Section \ref{reachable} presents a simplified temporal network version for such a cyber-social population with the proxy of campus WiFi accesses, where the reachability of such individual-level temporal networks affects the efficiency of information spreading. Section \ref{eventInteraction} defines the so-called event interaction to capture the concurrent interactive patterns among a group of individuals, and constructs the transmission graph to embed more temporal information, where both the vertex and edge dynamics of transmission graphs present rich temporal patterns. In Section \ref{temporalhub}, we introduce the outlier performance of the involved low-degree vertices' role in such temporal networks, which has been underestimated in the literature of static networks.  We define a temporal quantity as the participation activity potential to feature the role of individuals in a population, which achieves the prediction accuracy of ranking the individual centrality as high as 100\% with the verification of the WiFi data set. Finally, Section \ref{conclusion} concludes the whole chapter.

\section{Background}\label{background}

\begin{figure}[!t]
\centering
\includegraphics[width=3.5in]{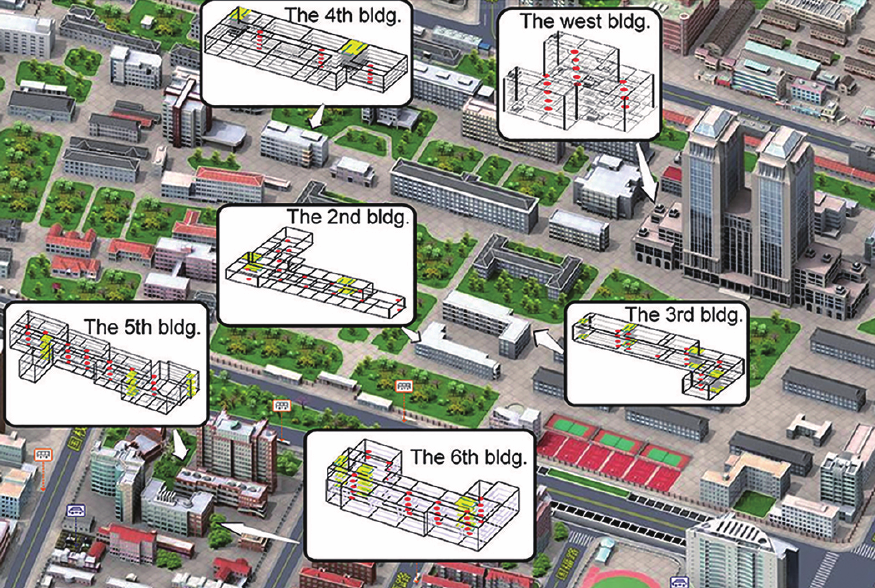}
\caption{The spatial distribution of wireless access points of six teaching buildings.The vertical view of `Handan Campus' of Fudan University, and the internal structures of six teaching buildings with the detailed deployments of wireless access points (red dots).}
\label{maps}
\end{figure}

Fudan University covers four campuses, and \emph{Handan Campus} is the oldest and also the largest campus, which accommodates around 75\% of students of the whole university. \emph{Handan Campus} is the first campus of Fudan university covered by WiFi signal, and in 2009 there have been 517 wireless access points (WAPs) distributed in all main buildings, which continuously provide dual-band (802.11g/n (2.4-GHz) and 802.11a (5-GHz)) wireless accessing services to all authorized campus members, \emph{i.e.}, students, teachers, staffs, and visiting scholars of Fudan University.

In the campus WiFi system, all WAPs share the same service set identifier (SSID), which guarantee that wireless devices/clients (e.g., laptops, pads, video game consoles, smart phones and digital audio players) can smoothly roam from one WAP to another.  Each WAP owns a unique IP address. When a client is roaming, the WiFi system automatically records the access logs without notifying the WiFi user. The assess logs of all authorized users are recorded in a database administrated by the Informatization Office of Fudan University. With the permission and assistance of the Informatization Office, we collected the campus WiFi users' access logs during the 2009-2010 fall semester (10/18/09-01/09/10) as the so-called data set of ``FudanWiFi09'', each piece of which contains four parts: the Media Access Control (MAC) address of the wireless device, the device's connecting and disconnecting time stamps, and the MAC address of the WAP that the device accesses to. Totally, the data set of ``FudanWiFi09'' contains the records of 22,050 WiFi devices' 423,422 behavioral trajectories in the period of nearly three months.

Since we target the proxy data of those wireless devices to feature the users' interactive behaviors, we only focus on the WiFi data recorded in the buildings which are open to public without safeguards for the WiFi users' devices, i.e., when a user leaves such a place, he/she should bring the device with himself/herself. Therefore, the data set of 18,715 individuals' 262,109 behavioral trajectories from six teaching buildings with the WiFi coverage have been employed in this work. Fig.\ref{maps} presents the spatial distribution of all 129 WAPs deployed in the six buildings from the vertical view of a part of \emph{Handan Campus} of Fudan University.

For simplicity, we assume that every WiFi device with the MAC address recorded in the WiFi system represents one WiFi user, neglecting the seldom cases that one person may use several WiFi devices at the same time, and several WiFi users may share the same WiFi device. Before transforming the access logs of WiFi data as the proxy of a large-scale population of WiFi users' interactions, we state our main assumption that the WiFi devices ``seeing'' the same WAP imply the co-located interactions among these devices' owners, which is motivated by the assumption of `geographic coincidences' proved in the work of Crandall and et al. in 2010 \cite{crandall2010inferring}. They defined a spatiotemporal co-occurrence between two Flickr users as an instance where they both took photos at approximately the same place and at approximately the same time, which infer the social ties in the population of humans.

Therefore, to verify the assumption of `geographic coincidences' in the campus WiFi system, the Euclidean distance between two WiFi users is critical. Due to the difficulty of locating the WiFi devices' indoor positions with a single WAP, we can not exactly calculate the Euclidean distance between two WiFi users in the campus. However, we can estimate the average Euclidean distances by Voronoi decomposition with the detailed distribution of WAPs shown in Fig. \ref{maps} and the student number in every classroom from `Fudan University Curriculum Schedule of the $1^{st}$ Semester in 2009-2010'. Since each building has several floors, and the building materials between floors can dramatically attenuate the wireless signals, which guarantee that the coverage regions of two WAPs on different floors do not overlap. On each floor, the wireless devices automatically connect to the closest WAP which offers strongest signals. Only when the WAP is overloaded, the device will switch to another adjacent WAP. In the campus WiFi system of Fudan University, each WAP can serve 50 users at most, and all the involved WAPs were not overloaded in the 3 months. Therefore, we decompose each floor by the distribution of deployed WAPs into the corresponding Voronoi tessellations. Given a Voronoi cell, we apply the Monte-Carlo method that randomly locating the persons in the classrooms of the cell to calculate the Euclidean distances of any two WiFi users, and estimate the average Euclidean distances and the standard deviations of any two WiFi users in every building, as summarized in Table \ref{table:euclidean}. We conclude that the average Euclidean distances along with their standard deviations do not change much in different buildings, indicating that the assumption of `geographic coincidences' is valid. Therefore, in such a close indoor circumstance, people have a high probability to directly communicate with each other or have some relationship via indirect interactions. In the following sections, we utilize the collected data of campus WiFi assess logs to explore categories of interactive patterns in a cyber-social population, which is illustrated with the population of WiFi users in Handan campus of Fudan University.

\begin{table}[!t]
\renewcommand{\arraystretch}{1.3}

\caption{The estimated average Euclidean distances and standard deviations between any two WiFi users.}
\label{table:euclidean}
\centering
\begin{tabular}{|c|c|c|c|}
\hline
\textbf{Building} & \textbf{No. of WAPs} & \textbf{$\langle d\rangle$(m)} & \textbf{$\delta$(m)} \\\hline
the $2^{nd}$ bldg. & 16 & 6.26 & 1.96\\\hline
the $3^{rd}$ bldg. & 21 & 7.14 & 0.96\\\hline
the $4^{th}$ bldg. & 15 & 7.86 & 1.71\\\hline
the $5^{th}$ bldg. & 28 & 6.98 & 1.37\\\hline
the $6^{th}$ bldg. & 33 & 7.11 & 1.59\\\hline
the west bldg. & 14 & 11.75 & 0.98\\\hline
\end{tabular}
\end{table}

\section{Pairwise interactive patterns of temporal contacts and reachability}\label{reachable}

\begin{figure}[!t]
\centering
\includegraphics[width=3.5in]{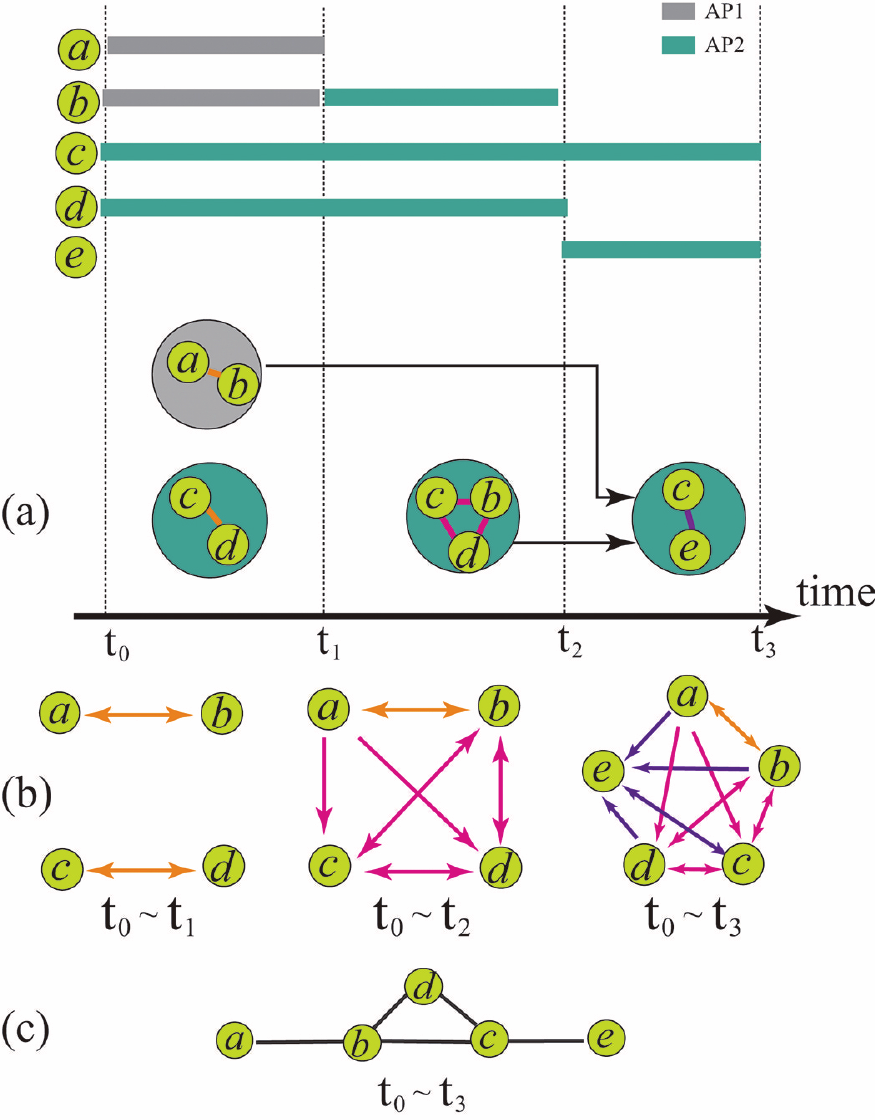}
\caption{The illustration of temporal contact network and its aggregated version. (a) WiFi accessing logs are translated into
the corresponding interaction events, where the black arrowed lines show the temporal contacts between users in different interaction events. (b) The
construction of three temporal contact networks during the intervals [t$_0$, t$_1$], [t$_0$, t$_2$], [t$_0$, t$_3$], respectively, where the edges are colored according to the latest time that they are contacted. (c) The construction of the aggregated contact network during the interval [t$_0$, t$_3$].}
\label{parta_definitions}
\end{figure}

We start from a simplified version of WiFi assess logs as shown in Fig. \ref{parta_definitions} (a), where five users ($a,b,c,d,e$) accessing to two different WAPs (AP 1 and AP 2) during the interval of $[t_0, t_3]$ are illustrated. Obviously, as time evolves, user $a$ interacts with user $b$ before user $b$ interacts with users $c$ and $d$. Since we assume that the WiFi devices ``seeing'' the same WAP infers the co-located interactions among these devices' owners, such `geographic coincidences'  indicate the temporal contacts among the involved WiFi users co-located in the same WAP. Therefore, a temporal contact network comes as a direct extension of contact networks with embedding the temporal information.

To construct a temporal contact network, we define ${e_i}=({V_i},{t_{i1}},{t_{i2}})$ as an interaction event, where ${V_i}$ denotes the set of users
connecting to the same WAP simultaneously, and any user $v\in {V_i}$ interacts with other users in ${V_i}$  during the
interval from ${t_{i1}}$ to ${t_{i2}}$. We define a temporal contact (TC): user $i$ makes a temporal contact with user $j$ if there exists a
sequence of interaction events with non-decreasing time between them. Therefore, a temporal contact network is constructed at the level of individual users, \emph{i.e.}, define a temporal contact network $\mathcal{G}=\{\mathcal{V},\mathcal{E}\}$ with $\mathcal{V}$ the set of vertices (WiFi users) and $\mathcal{E}$ the collections of all temporal contacts. Note that $\mathcal{E}=\mathcal{E_{\rightarrow}}\bigcup\mathcal{E_{\rightleftharpoons}}$ is the union set of directed edges $\mathcal{E_{\rightarrow}}$ and bidirectional edges $\mathcal{E_{\rightleftharpoons}}$, which stand for the time-respecting paths and pairwise interactions, respectively. Given an observed interval, as shown in Fig. \ref{parta_definitions} (b), there exist two time-respecting paths: one path is from user $a$ to user $c$, and the other is from user $a$ to user $d$. Besides, there exist two pairwise interactions between two pairs of users $b$, $c$, and users $b$, $d$.
Therefore, we generate three different temporal contact networks in different intervals $[t_0, t_1], [t_0, t_2], [t_0, t_3]$, respectively. As a comparison, we aggregate all interaction events in the interval of $[t_0, t_3]$, and generate an aggregated contact network as shown in Fig. \ref{parta_definitions} (c), which clearly tells the difference between two different contact networks even in the same interval $[t_0, t_3]$.

\begin{figure}[!t]
\centering
\includegraphics[width=3.5in]{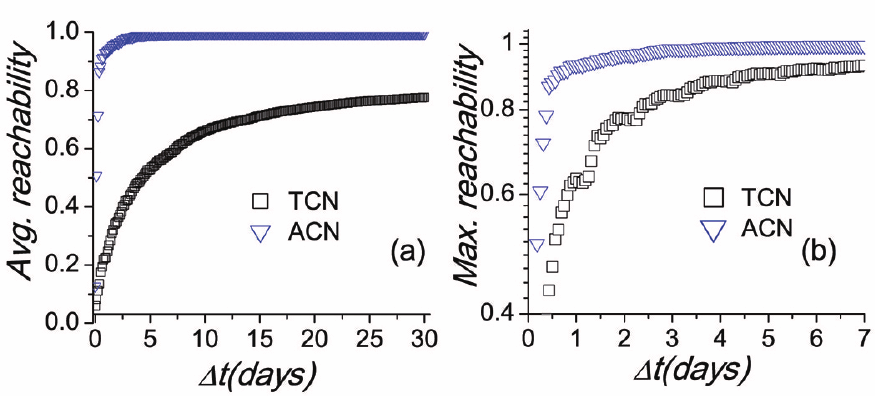}
\caption{Network reachability. (a) The average network reachability of the aggregated contact networks (ACNs) and temporal contact networks (TCNs) with the time length $\Delta$t=30 days. (b) The maximum network reachability of the ACNs and TCNs  with the time length $\Delta$t=7 days.}
\label{reachability}
\end{figure}

Such different connectivity patterns between temporal contact networks and aggregated contact networks are also reflected in the basic property of network connectivity: reachability. Traditionally, in the case of aggregated (static) networks, since the edges are continuous active, the size of the component, $N_{i}$,  that a given vertex $i$ belongs to, indicates an upper bound of the number of vertices influenced by vertex $i$. Therefore, $N_{i}$ characterizes the reachability of vertex $i$ in the static contact network aggregated within a given time interval. In the case of temporal (contact) networks, however, the edges are no longer continuously active, whose active durations and time stamps dominate the network transitivity. Since a temporal contact network defined here is rather simplified, which only keeps the contact order of time stamps, the reachability of vertex $i$ in a temporal (contact) network is the number of vertices that can be temporally influenced by  $i$ during the time length $\Delta t$ \cite{holme2005network}. More specifically, the reachability of vertex $i$ of a temporal contact network equals to the number
of elements in the set $\{ {\phi _{i,j}}(t_2)\mid {\phi _{i,j}}(t_2) > {t_1}, \ j \in V\}$, where $\phi_{i,j}(t)$ is the time of the inception of the latest temporal contact from $i$ to $j$ before $t$. The average (maximum) network reachability is the mean (maximum) value of all vertices' reachability.

Fig. \ref{reachability} compares the network reachability (normalized by the network size) between temporal contact networks and aggregated contact networks with the same time length. Both the average and maximum values tell that the aggregated contact networks' reachability is much larger than those of temporal contact networks, especially when the time length $\Delta t\rightarrow 0$. Besides, the reachability of aggregated contact networks quickly attains its saturation with increasing the time length $\Delta t$, while the reachability of temporal contact networks saturates much slower. Since the saturated value of the average reachability of temporal contact networks is much smaller than that of aggregated contact networks, the temporal dimension provides a more precise upper bound to the reachability of a network, which may avoid the overestimation of the information efficiency over such networks.

\begin{figure}[!t]
\centering
\includegraphics[width=3.5in]{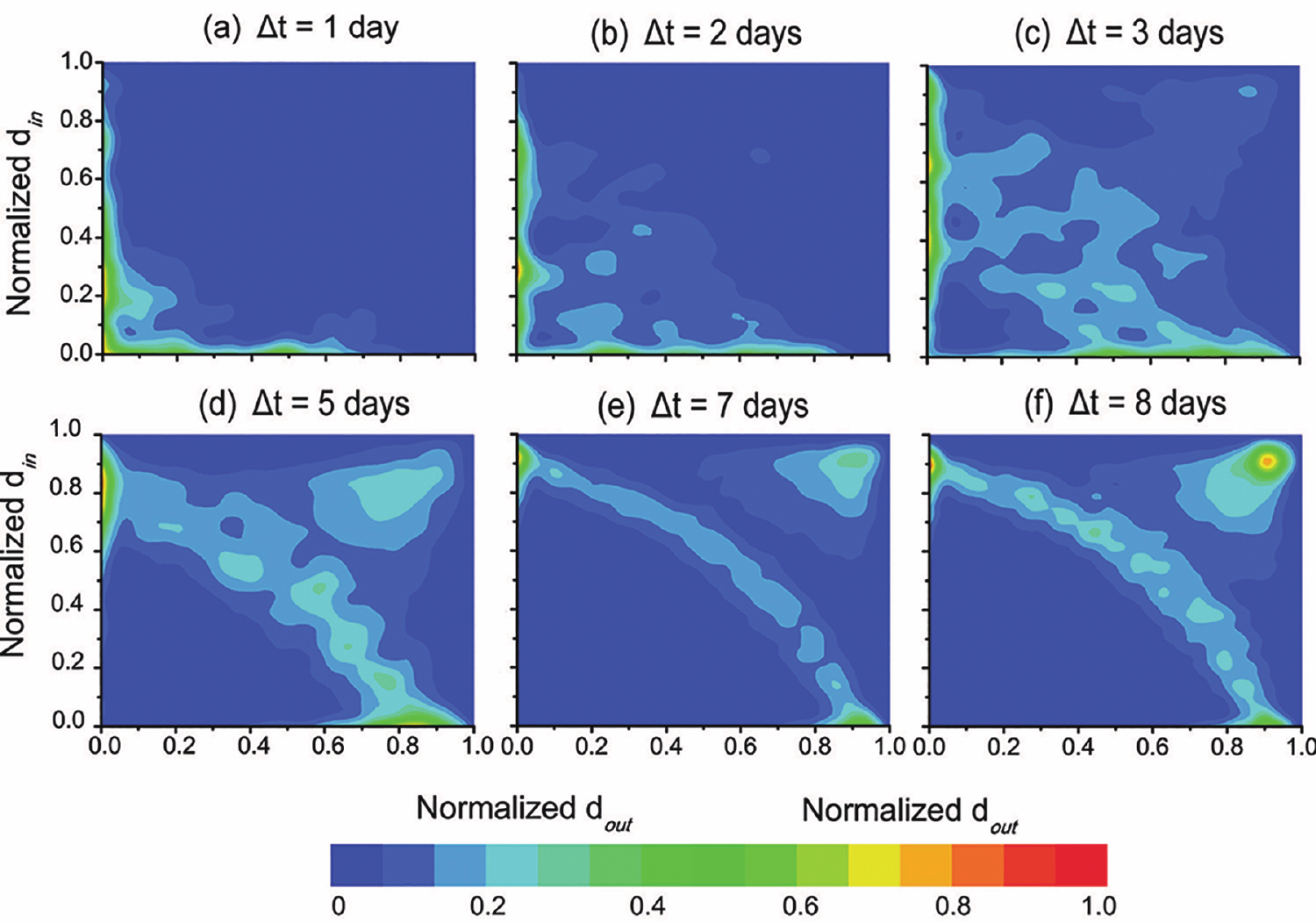}
\caption{The correlation of the temporal out-degrees and in-degrees. The joint probability distribution of the temporal out-degrees d$_{out}$ and in-degrees d$_{in}$ with the given time length $\Delta$t as (a) 1 day, (b) 2 days, (c) 3 days, (d) 5 days, (e) 7 days, (f) 8 days. When the time length $\Delta$t is larger than 1 week (7 days), a set of data points always present at the right-up corner.}
\label{vertices}
\end{figure}

In a temporal contact network, the out-degree $d_{out,i}$ of vertex $i$ quantifies the number of receptors temporally affected by $i$, while the in-degree $d_{in,i}$ specifies the number of its potential inciters. In an information spreading process on a temporal (contact) network, for example, except the source/destination vertices as well as those leaf vertices, all other vertices in the network play both roles of receiving and forwarding information in the process. Therefore, we put more attention to the correlated distribution of $d_{out,i}$ and $d_{in,i}$ of all vertices in the temporal contact networks with different time lengths $\Delta t$. Fig.\ref{vertices} presents the joint probability distributions $C^{\Delta t}(d_{out},d_{in})$ of out-degrees and in-degrees with different time lengths $\Delta t=1,2,3,5,7,8$ days, respectively. Each joint probability distribution is averaged by the generated networks from the data set  with a given $\Delta t$. In the case of the smallest daily time length, \emph{i.e.}, $\Delta t=1$ day, Fig.\ref{vertices}(a) shows that almost all data points reside in the lower triangular matrix, and many data points are even sticking on the axes. That is to say, in a short observation period, \emph{e.g.}, 1 day, only a very limited number of users keep frequently online during the whole day, while most individuals act according to their curriculum schedules, which seldom cover the whole day. With the increase of time length $\Delta t=2,3,5$ days, there are more and more vertices emerging in the upper region of two axes (Figs.\ref{vertices}(b)-(d)), which tell that the forward and backward temporal influence of each user increases as time proceeds. When the time length is more than one week (Figs. \ref{vertices}(e) and (f)), there are two evidently nontrivial clusters of temporal hubs: one contains the vertices along the diagonal, and the other contains the vertices anchoring in the upper right corner. In the former cluster of hubs, the users balance to receive and forward the influence from other users. While those users presented in the upper right corner are more important due to their role of relay hubs, which are critical to facilitate spreading processes on temporal networks. Obviously, the weekly cycle of the curriculum schedules of all campus WiFi users accelerates the emergence of such temporal patterns of relay hubs.

\section{Concurrent interactive patterns of event interactions and temporal transmission graphs}\label{eventInteraction}

Temporal contact networks as a category of ``reachability graph'' are generated based on the ``time-respecting'' paths and pairwise interactions of temporal contacts, which focus on the temporal interactive sequence between a pair of individuals while neglect the concurrent interactive dynamics of more involved individuals. For example, recall Fig. \ref{parta_definitions} (b), the three users $b,c,d$ concurrently assess `AP 2' in the overlapped interval $[t_1, t_2]$, while this concurrent interaction among three users has to be simplified to two pairwise interactions ($bd$ and $bc$) with the definition of temporal contacts. To characterize such concurrent interactions, we extend the temporal contacts at the level of individual users to a group of users with the newly defined `event interactions' and `transmission graphs' to embed more temporal information.

\begin{figure}[!t]
\centering
\includegraphics[width=3.5in]{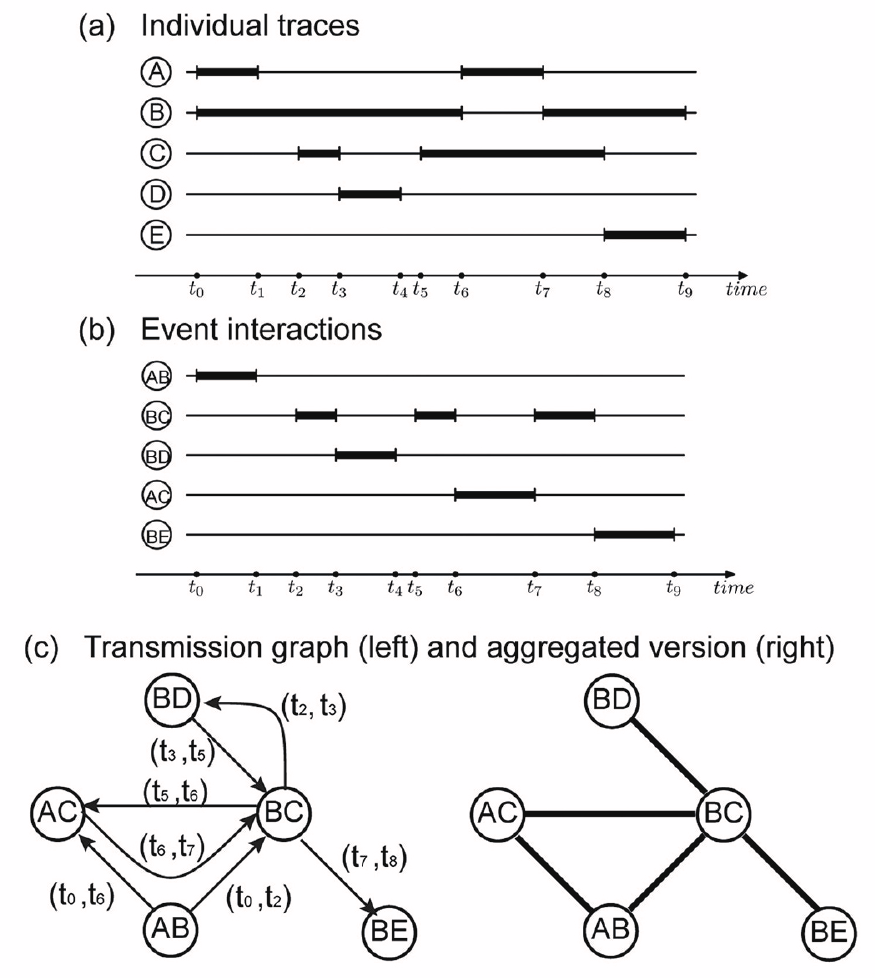}
\caption{The illustration of event interactions and transmission graph. (a)WiFi access logs as the proxy of the users' behavioral trajectories. The bold lines pertain to their online durations. (b) Each bold line possesses an exclusive time interval, where the corresponding individuals are assembled into a contact clique. (c) In a transmission graph, the vertices are event interactions (EIs), and the edges between two EIs are the transmission paths defined by the three rules. The aggregated version is derived from the transmission graph, where no multiple edges and time labels are allowed.}
\label{schematic}
\end{figure}

Fig. \ref{schematic} (a) gives an instance of five WiFi users' access logs as the corresponding individual behavioral traces, where the bold lines pertain to their online durations, \emph{i.e.}, the five users stay at the coverage of the same WAP during [$t_0,t_9$], and only these five users access to this WAP during the interval. As we assume that the WiFi devices ``seeing'' the same WAP infers the geographic co-located interactions among these devices' owners (\emph{i.e.}, co-locating at the same time in the same small region), the five users participate several access events when different users access and disconnect the WAP at different time stamps. Therefore, we define an event interaction as the concurrent interaction of multiple users accessing to the same WAP within a given interval, \emph{e.g.},  EI $E_{AB}^{t_0}$ characterizes the concurrent interactions of users \emph{A} and \emph{B} at the beginning time $t_0$. Fig. \ref{schematic} (b) illustrates the process of translating the five users' individual behavioral trajectories into event interactions (EIs). Since a user only involves in one unique event interaction at any time, all illustrated events in Fig. \ref{schematic} (b) are involved with two users for simplicity, and an event interaction with more users involved is still valid, which is reflected in the later defined quantity of event size.

Define an event interaction (EI) as a vertex, and two vertices (event interactions) are connected in a `Transmission Graph' with the following three defined rules:

\begin{enumerate}
  \item At least one same user exists in both the source and sink EIs.
  \item In the time sequence, the beginning time of the source EI is the closest one prior to the beginning time of the sink EI.
  \item Given a sink EI, when there exist several source EIs, the set of shared users in a pair of source EI and sink EI does not intersect with those of other pairs of source and sink EIs.
\end{enumerate}

The first two rules follow the principle of temporal nearest adjacency, and the third rule follows the principle of minimum number of transmission edges (paths) generated. Therefore, we construct a transmission graph as illustrated in Fig. \ref{schematic} (c) with the above three rules, yielding five vertices (event interactions of Fig. \ref{schematic} (b)) and seven directed edges, where the time stamps such as $t_0, t_6$ over the directed edge between vertex $AB$ and vertex $AC$ stand for the beginning time stamps of the two (source and sink) event interactions. Also, we aggregate all event interactions following the three rules of transmission edges with the same data set to generate an aggregated transmission graph, which is illustrated in Fig. \ref{schematic} (d).

To quantify the temporal information of event interactions (EIs) and the transmission graph (TG), we further define the following quantities:

\begin{itemize}
  \item The size, \emph{s}, of a given EI is defined as the number of involved individuals.
  \item Given an EI active at the epoch of [$t^{begin},t^{end}$], its active duration is defined as $\Delta t^{EI}=t^{end}-t^{begin}$.
  \item Given a pair of source and sink EIs, the time-stamped label in the directed edge is $(t_{source},t_{sink})$. The transmission duration $\delta$ is defined as the interval between the beginning time of the source and sink EIs, \emph{i.e.}, $\delta = t_{sink}-t_{source}$.
\end{itemize}

\begin{figure}[!t]
\centering
\includegraphics[width=3.5in]{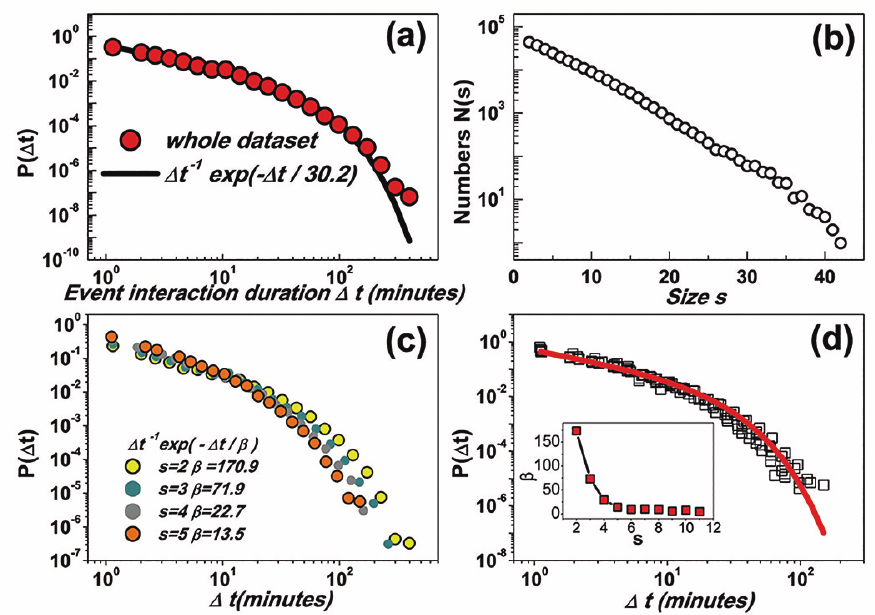}
\caption{Temporal dynamics of event interactions. (a) The probability distribution of event interactions' active durations $\Delta$t$^{EI}_{all}$. (b) The size distribution of event interactions. (c) The probability distributions of event interactions' active durations $\Delta$t$^{EI}_{s}$ with the given size s = 2,3,4,5, respectively. (d) The probability distributions of event interactions' active durations $\Delta$t$^{EI}_{s}$ with the given size s=5,6,...11, respectively. The red bold line is the distribution with s=5. The inset tells the dependence between size s, and the exponent of exponential cutoff $\beta$.}
\label{dynei}
\end{figure}

Since the vertices of a transmission graph are the event interactions, we first present the vertex dynamics of transmission graphs to unveil the statistical characteristics of concurrent interactions among WiFi users. As shown in Fig. \ref{dynei} (a), the probability distribution of all event interactions' active durations collected during the three months falls into a truncated power law, whose exponent of the power-law part approximates to 1. This feature indicates that the long-lasting concurrent interactions can hardly survive due to those newly emergent events with short active durations, while the non-Poission distribution of concurrent interactive durations tells that WiFi users do not randomly break their interactions with other users. On the other hand, the number of event interactions with a given size is exponentially distributed as shown in Fig. \ref{dynei} (b), \emph{i.e.}, the users randomly co-locate with each other at the same time.

Note that such an exponential distribution also implies that the large-size event interactions are rare, and we may conjecture there exists some interdependence between the probability distribution of event interactions' active durations $\Delta t^{EI}$ and their sizes \emph{s}, which, however, does not exist as shown in Figs. \ref{dynei} (c)-(d). Although the probability distributions of active durations $\Delta t^{EI}_{s}$ with size $s=2,3,4,5$ keep the truncated power laws, whose power-law parts still obey the exponents close to 1, we observe that their exponential cutoffs gradually decay with the increased event size from 2 to 5 (Fig. \ref{dynei}(c)). Furthermore, with the increased event size from 6 to 11 as shown in Fig. \ref{dynei}(d), the probability distributions of the active durations with size $s\geq 5$ present the same shape $\Delta t^{-\gamma}\exp(-\frac{\Delta t}{\beta})(\gamma\approx 1)$. At the same time, the inset of Fig. \ref{dynei}(d) tells that the exponent of exponential cutoff $\beta$ decreases with the growth of size \emph{s} from 2 to 5, which keeps invariant with increasing \emph{s} from 5 to 11. Therefore, we conclude that the active durations of event interactions own the \emph{`size free'} feature, which, counter-intuitively, is independent from the event sizes.

\begin{figure}[!t]
\centering
\includegraphics[width=3.5in]{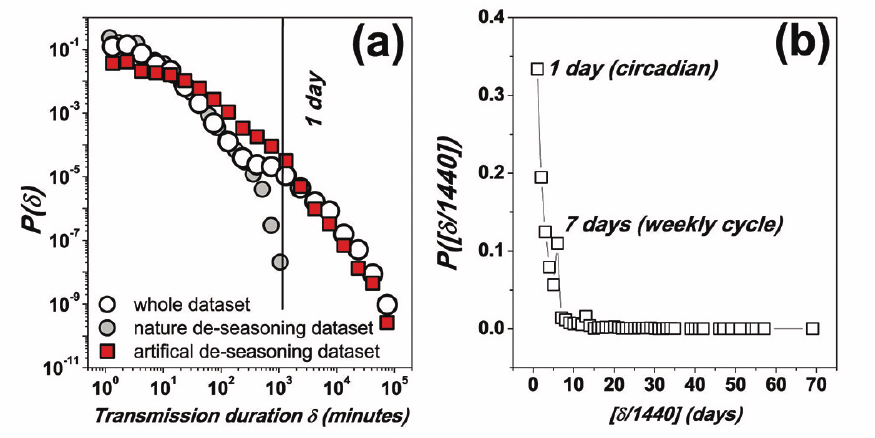}
\caption{Temporal dynamics of transmission durations. (a) The probability distribution of transmission durations, $\delta$, from the whole dataset (white circle), and `filtered' by the natural de-seasoning method (gray circle) and the artificial de-seasoning method (red square).(b) The probability distribution of integral days spent by transmission duration $[\frac{\delta}{1440}]$ (days).}
\label{dyntg}
\end{figure}

On the edge dynamics of transmission graphs, we focus on the transmission durations to uncover the temporal relations between concurrent interactions. As shown in Fig. \ref{dyntg} (a), the probability distribution of transmission durations obeys a bifold power-law with two turning points: the first turning point is at around 120 minutes, \emph{i.e.}, the length of two teaching courses; the other turning point is at around 24 hours, which indicates the existence of daily circadian rhythms. To test the existence of periodic rhythms, we calculate the integral days spent by each $\delta$ ([$\frac{\delta}{1440}$]), as shown in Fig. \ref{dyntg}(b), where there are two peaks at the 1st day and the 7th day. That is to say,  the daily burst behavior of `adjacent' event interactions evolves with the weekly rhythm. We further employ two (natural and artificial) de-seasoning methods to remove the circadian and weekly rhythms. In the natural de-seasoning method, we conserve the source and sink event interactions taking place in the same day. In the artificial de-seasoning method, we randomly select two event interactions and exchange their beginning active time (their active durations are ignored), and check all individuals to ensure this round of time-shuffle operation does not generate the event interactions with the same involved individual which have the same beginning time. As shown in Fig. \ref{dyntg}(a), the probability distributions of the transmission durations `filtered' by two de-seasoning methods both fall into the truncated power-laws, and we witness that such burst behaviors of populations take place \emph{independently} on the daily and weekly rhythms.

\section{Temporal degrees and hubs: ranking and prediction}\label{temporalhub}

In Sec. \ref{reachability}, those users located in the upper right corner of Fig. \ref{vertices} (f), i.e., users with both large in-degrees and out-degrees, function as the relay hubs of temporal contact networks, which emerge with the weekly rhythms of WiFi users' curriculum schedules. In this section, we further explore the temporal hubs with more temporal information embedded into event interactions and transmission graphs as defined in Sec. \ref{eventInteraction}. When neglecting temporal information, traditionally, a static contact network with the same data set of WiFi campus assess logs is generated, where the vertices are WiFi users, and an edge connects two vertices if the two WiFi users have at least one round of interaction (\emph{i.e.}, geographic coincidence) during the whole 3 months.  As a comparison, we aggregate all transmission edges of Fig. \ref{schematic} (c) (left) to generate an aggregated transmission graph, as illustrated in Fig. \ref{schematic} (c) (right).

\begin{figure*}[!t]
\centering
\includegraphics[width=5in]{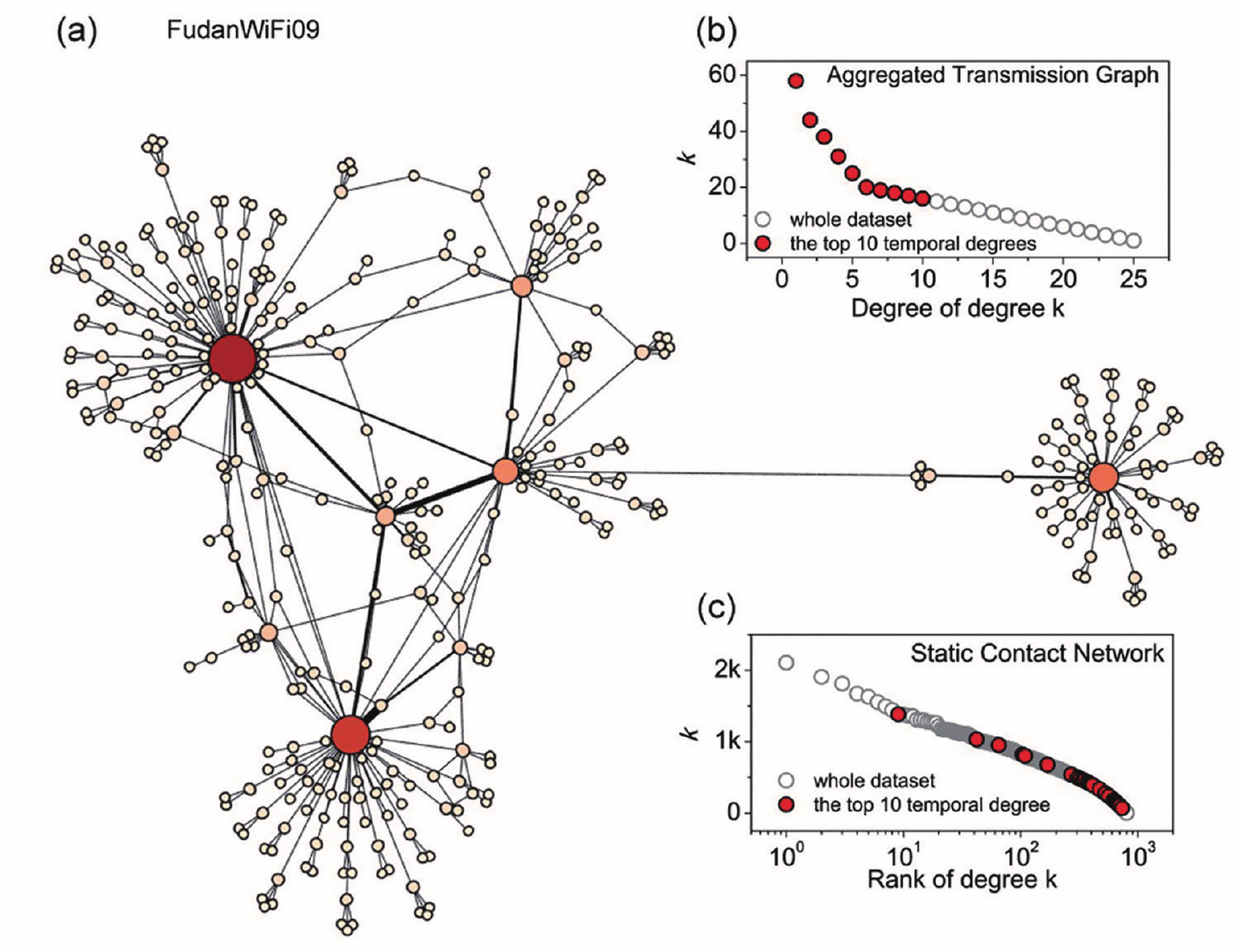}
\caption{Outliers: temporal hubs in the aggregated transmission graph versus leaf vertices in the static contact network. (a) A small sample of the aggregated transmission graph consisting of several hubs that WiFi users involved in. (b) The degrees versus their ranks of all WiFi users in the aggregated transmission graph. The red circles represent the WiFi users who have the top 10 temporal degrees.(c) The degrees versus their ranks of all WiFi users in the static contact network with the same WiFi data set. The red circles represent the same WiFi users in (b).}
\label{leafhub}
\end{figure*}

Both temporal and static aggregated networks present some differences. The static contact network defines the vertices at the individual level of WiFi users, which is topological homogeneous with the degrees' coefficient of variation (CV) as $1.59$. While the aggregated transmission graph defines the vertices at the level of event interactions, which concurrently involve a group of WiFi users, therefore, it is more heterogeneous with the degrees' coefficient of variation as $7.13\gg 1$. Fig. \ref{leafhub} (a) visualizes a small sample of the aggregated transmission graph. Notice the degree of event interactions (vertices) in the aggregated tranmission graph is the sum of the out-degree and in-degree in the corresponding transmission graph, we define the degree of an event interaction as the degree of the event interaction's involved users. Generally, a (WiFi) user may participate in several event interactions with different degrees, we define the temporal degree of a user as the maximal degree of all the involved event interactions. Therefore, we rank the temporal degrees of WiFi users as shown in Fig. \ref{leafhub}(b), where the top 10 users with the largest temporal degrees are identified with red circles. Interestingly, these temporal-hub users are not so `attractive' in the static contact network, which, on the contrary,  are almost the low-degree vertices as identified with red circles in Fig. \ref{leafhub}(c). That is to say, many low-degree vertices, which generally were  neglected in the literature of static contact networks, may participate to play the dominant role as `hubs' in the version of temporal networks. Such difference also highlights the temporal significance of concurrent interactions to characterize a typical cyber-social population with the society example of a WiFi campus, and the outlier nontrivial performance has also been verified in other public data sets \cite{zhang2013temporal}.

Note that we can not directly identify the WiFi users who function as temporal hubs from the traditional version of static contact networks, as we have visualized  that the low-degree individuals in the static human contact networks gather together to form temporal `hubs' in the aggregated transmission graphs. Therefore, to identify (and even predict) such important users who have high temporal degrees in the WiFi campus, a natural means is constructing such an aggregated transmission graph before ranking all involved users with their temporal degrees, which, however, is a task with very highly computational load. For example, given a temporal data set, which yields $M$ event interactions with time stamps, we construct a transmission graph having $H$ vertices and $P$ edges, whose computational complexity in the worst case is $O(M^2)$. While the process of aggregating such a transmission graph requires the computational load of $O(H*P)$. Therefore, with the WiFi assess data addressed in this chapter, we totally generate $M=260,925$ event interactions and a transmission graph having $H=248,663$ vertices and $P=260,513$ edges, whose computational load is as high as $O(1.32*10^{11})$ to generate the aggregated transmission graph, excluding the search cost to rank temporal degrees. Obviously, a new and efficient method to rank and predict the temporal hubs without constructing such a transmission graph is desirable and challenging, and we give a satisfactory solution as follows.

\begin{figure}[!t]
\centering
\includegraphics[width=3.5in]{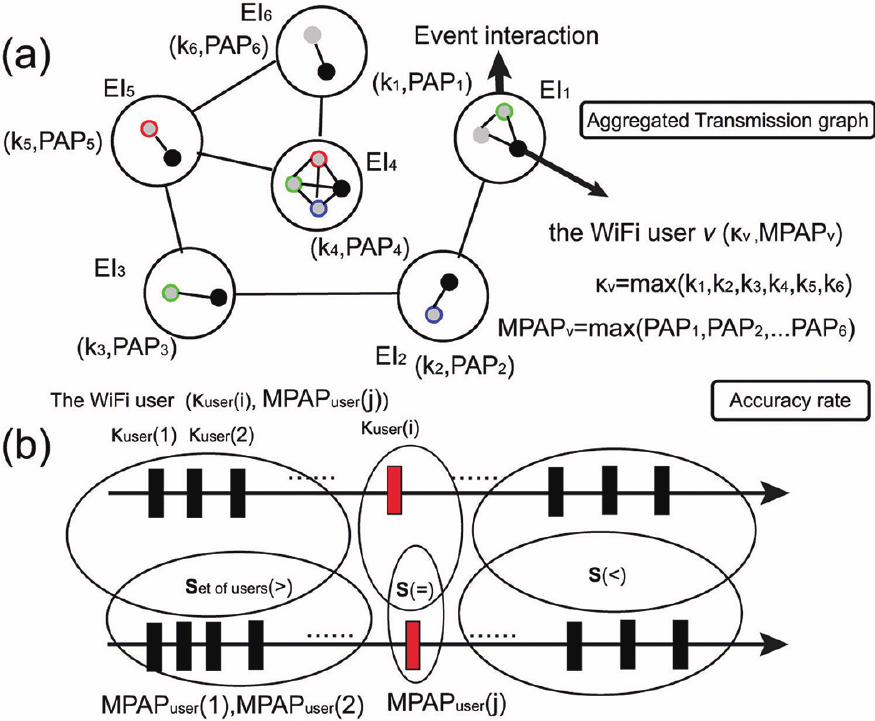}
\caption{The illustration of predicting temporal hubs. (a) In an aggregated transmission graph, each vertex is an event interaction (EI), and each EI contains several Wi-Fi users. The temporal degree, $\kappa$, of a Wi-Fi user is the maximum degree of the involved EIs, and the corresponding maximum participation activity potential (MPAP) is the maximum value of all participation activity potentials (PAPs) of the involved EIs. (b) The temporal degree, $\kappa$, and MPAP are both in descending order. Given a Wi-Fi user, its accuracy rate is the ratio of the other Wi-Fi users who are in the same positions predicted by both $\kappa$ and MPAP.}
\label{illustration_predicating}
\end{figure}

As shown in Fig. \ref{illustration_predicating} (a), the vertices of an aggregated transmission graph are event interactions(EIs), which contain several WiFi users followed by the definitions of transmission graph in Section \ref{eventInteraction}. The active duration of an event interaction, $e_i$, is defined as $\Delta t^{EI}(e_i)$. Given a finite length of time period $\Delta T$ and an event interaction, $e_i$, the total active (WiFi users') number of the event interaction is $n(e_i)$, and we define the total active duration of the event interaction as $$\Delta t^{EI}_{sum}(e_i)=\sum_{m=1}^{n(e_i)}\Delta t^{EI}_m(e_i).$$

Given the WiFi user $j$ participates event interaction $e_i$, we define the \emph{participation activity potential(PAP)} of user $j$ involving event interaction $e_i$ as:

\begin{equation}\label{PAP}
    PAP(j,e_i)=(\Delta t^{EI}_{sum}(e_i))^{\alpha}(n(e_i))^{1-\alpha},\alpha\in[0,1]
\end{equation}

As we know, a WiFi user may involve several different event interactions, therefore, given the set of event interactions involved by WiFi user $j$ as $\Gamma(j)=\{e_i|j\in e_i\}$, we define the \emph{maximum participation activity potential (MPAP)} of user $j$ as:

\begin{equation}\label{MPAP}
    MPAP(j)=\max_{e_i\in\Gamma(j)}(PAP(j,e_i))
\end{equation}

\begin{figure*}[!t]
\centering
\includegraphics[width=6in]{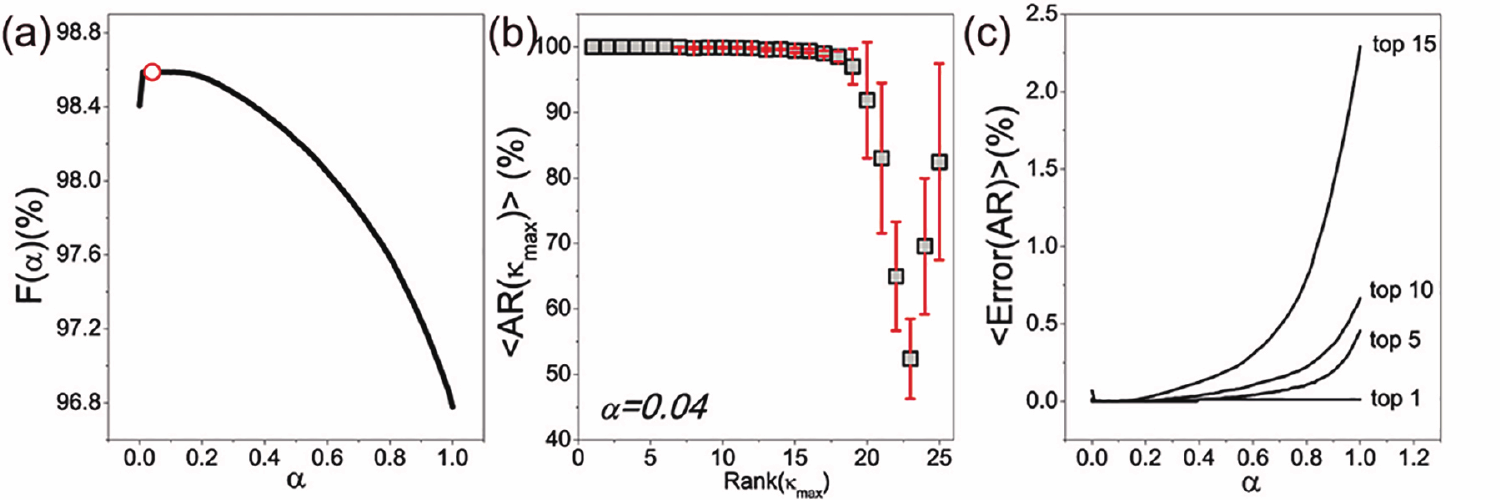}
\caption{Predicting temporal hubs of a transmission graph. (a) The weighted accuracy rate F($\alpha$) shows that the accuracy rate of predicating temporal hubs achieves more than 96\% when randomly selecting a value of $\alpha$ from [0,1]. The red dot represents that the maximum weighted accuracy rate can achieve 98.6\% with a given optimal nonlinear parameter $\alpha$=0.04. (b) The average accuracy rate $\langle$AR($\kappa_{max}$)$\rangle$ of predicating temporal hubs versus the rank of their temporal degrees $\kappa_{max}$ with the optimal $\alpha$=0.04. The red bar represents the standard deviation $\delta$(AR($\kappa_{max}$)) of the accuracy rate. (c) The average error between the accuracy rate given a random nonlinear parameter, $\alpha$, and the optimal $\alpha$=0.04 to predict top 1, 5, 10, 15 Wi-Fi users.}
\label{algorithm}
\end{figure*}

Different from temporal degrees to characterize WiFi users' temporal role, which require a heavy computation cost as we have discussed, the new proposed quantity of maximum participation activity potential (MPAP) does not need any preliminary information from the aggregated transmission networks.  For instance, in Fig. \ref{illustration_predicating}(a), WiFi user $v$ involves in six event interactions with the corresponding degree $k_i$ and the participation activity potential $PAP_i$, where $i=1,2,...6$. The temporal degree of user $v$ is the maximal degree of the involved six event interactions, $\kappa(j)=\max_{i=1,...,6} k_i$, which is dependent on constructing such an aggregated transmission graph. To calculate the maximum participation activity potential of user $j$, the maximum value of all PAPs of the involved event interactions, $MPAP(j)=\max_{i=1,...,6}PAP_i$, is directly calculated from the information of each event interaction itself.

Therefore, we propose the maximum participation activity potential (MPAP) of a user as the candidate quantity of `temporal degrees', with which we rank all WiFi users and predict the temporal hubs in the campus population. As shown in Fig. \ref{illustration_predicating}(b), we list all temporal degrees and all MPAPs of the WiFi users in the descending order, where $`\kappa_{user}(1)>\kappa_{user}(2)>...'$ and $`MPAP_{user}(1)>MPAP_{user}(2)>...'$. Given a WiFi user with temporal degree $\kappa_{user}(i)$, all the other WiFi users in the population are divided into the following three sets based on temporal degree $\kappa_{user}(i)$:

\begin{itemize}
  \item set $S_{\kappa}(>)$: the users whose temporal degrees $\kappa > \kappa_{user}(i)$.
  \item set $S_{\kappa}(=)$: the users whose temporal degrees $\kappa = \kappa_{user}(i)$.
  \item set $S_{\kappa}(<)$: the users whose temporal degrees $\kappa < \kappa_{user}(i)$.
\end{itemize}

Similarly, we divide all the same users into another group of three sets based on $MPAP_{user}(j)$:

\begin{itemize}
    \item set $S_{MPAP}(>)$: the users whose $MPAP > MPAP_{user}(j)$.
    \item set $S_{MPAP}(=)$: the users whose $MPAP = MPAP_{user}(j)$.
    \item set $S_{MPAP}(<)$: the users whose $MPAP < MPAP_{user}(j)$.
\end{itemize}

Finally, we find the part of WiFi users who present in the following sets as shown in Fig. \ref{illustration_predicating}(b):

\begin{itemize}
    \item the set of $S(>)$: the users present in $S_{\kappa}(>)\bigcap S_{MPAP}(>)$.
    \item the set of $S(=)$: the users present in $S_{\kappa}(=)\bigcap S_{MPAP}(=)$.
    \item the set of $S(<)$: the users present in $S_{\kappa}(<)\bigcap S_{MPAP}(<)$.
\end{itemize}

Therefore, the users ranked by their temporal degrees and the maximum participation activity potential in the same positions describe the accuracy rate predicted by the maximum participation activity potential defined as follows:

$$AR(v)=\frac{|S(>)|+|S(=)|+|S(<)|}{n-1}$$

In more detail, given a rank of the temporal degree $r(\kappa)$, we define the average predicative accuracy rate and its standard deviation as

\begin{equation}\label{MAR}
\langle AR(r_{\kappa})\rangle=\frac{\sum_{r_{\kappa}(v)=r}AR(v)}{|\{v|r_{\kappa}(v)=r\}|}
\end{equation}

\begin{equation}\label{SAR}
\delta(AR(r_{\kappa}))=\sqrt{\frac{\sum_{r_{\kappa}(v)=r}(AR(v)-\langle AR(r_{\kappa})\rangle)^2}{|\{v|r_{\kappa}(v)=r\}|}}
\end{equation}

Note that in the definition of participation activity potential $PAP$, Eq. (\ref{PAP}), an appropriate value of parameter $\alpha\in [0,1]$ influences the accuracy rate. To find the optimal $\alpha$ and achieve the highest prediction accuracy, we further define the following function $F(\alpha)$:

$$F(\alpha)=\frac{\sum\kappa*\langle AR(\kappa)\rangle}{\sum\kappa}.$$

As shown in Fig. \ref{algorithm} (a), the maximum accuracy ratio is $F(\alpha)=98.6\%$ with the corresponding $\alpha=0.04$ as the optimum, and a random selection of $\alpha\in(0,1)$ achieves the accuracy rate higher than 96\%. Substituting the optimum of $\alpha=0.04$ to the participation activity potential $PAP$ defined in Eq. (\ref{PAP}), the average accuracy rate (Eq. (\ref{MAR})) predicts the \emph{top 15} temporal hubs by the maximum participation activity potential $MPAP$ as high as 100\% accuracy rate (Fig. \ref{algorithm}(b)). Furthermore, to predict the different cases, for example, of the top 1, 5, 10, and 15 temporal hubs, the performance difference between a randomly selected $\alpha\in (0,1)$ and the optimal $\alpha=0.04$ is as less as $2.5\%$ at most, which is visualized in Fig.\ref{algorithm}(c). Note that such an optimal $\alpha=0.04$ is not universal, whose value is case by case generally \cite{zhang2012characterizing}. Therefore, a randomly selected $\alpha\in(0,1)$ may achieve a sub-optimal performance of the prediction in practice.

\section{Conclusion}\label{conclusion}

To summarize this chapter, with the proxy data of WiFi assess logs collected in a campus of Fudan University, we have presented the series of temporal interactive patterns of the large-scale population of WiFi users as a representative example of cyber-social populations in the networking era of today. The efforts from the viewpoint of temporal networks offer sufficiently powerful supports to discover human interactions, as we illustrated in this chapter, from both pairwise and event interactions defined at the levels of individual users and a group of concurrent users, respectively. Not limited to the rich temporal patterns of the WiFi-interactive population, the outlier performance of the temporal hubs along with the roles of leaf connectivity status also sheds more light to the significance of temporal networks, which, in return, help to rank and predict the important active users in the temporal framework.

Nevertheless, the understanding of network science from the view points of temporal networks as well as human population dynamics still leaves many topics to explore, which is far beyond the previous addressed issues such as time-varying, time-evolving, and time-switching networks. Especially, new concepts from not only the network topological connectivity but also dynamical processes and the collective performance need to bridge the gap between the traditional (static/aggregated) version and the now focused temporal dimension, for example but not limited to temporal degree, temporal path in previous sections, and temporal clustering coefficient \cite{Cui2013cct}, temporal betweenness\cite{Pfitzner2013bpq}, in and out-components\cite{Konschake2013rio}, temporal small world\cite{tang2010swb}, temporal motif \cite{kovanen2011temporal,kovanen2013temporal,zhang2013exploring}, and etc.

Return to the key question in the beginning of this chapter. The modeling of such fascinating and yet temporal complex features of a network very recently have reported several models as a combination of human dynamics and evolving networks \cite{Barrat2013mtn,Jo2011ebc,Min2009wtd,Perra2012adm}, and the dynamical processes \cite{Lee2012etn,rocha2013bursts,Machens2013idm} as well as the collective behaviors \cite{Perra2012rws,Starnini2013mhd} and even the structural controllability \cite{Pan2013sctn} over such temporal networks have newly attracted more broad attention. When finishing this chapter, we also notice the book entitled ``Temporal networks" \cite{Holme2013Book} which covers a more extensive scope of temporal networks for the interested readers' further reference, while, on the other hand, this chapter synthesizes this new emergent branch with a detailed WiFi-interactive social population.

\section*{Acknowledgment}
Xiang Li and Yi-Qing Zhang were grateful to the Informatization Office of Fudan University for the WiFi Data collection, and the Archives of Fudan University for providing the blueprint of all six teaching buildings. This work was partly supported by the National Key Basic Research and Development Program (No.2010CB731403), the National Natural Science Foundation (No.61273223), the Research Fund for the Doctoral Program of Higher Education (No.20120071110029) and the Key Project of National Social Science Fund (No. 12\&ZD18) of China.\protect\\

\ifCLASSOPTIONcaptionsoff
  \newpage
\fi

%

\begin{IEEEbiographynophoto}{Xiang Li}
(M'05-SM'08) received the B.S. and Ph.D. degrees in control theory and control engineering from Nankai University, China, in 1997 and 2002, respectively, and he currently is a Professor and Head of the Electronic Engineering Department, Fudan University. Prof. Xiang Li serves and served as Associate Editor of IEEE Transactions on Circuits and Systems-I: Regular Papers, IEEE Circuits and Systems Society Newsletters, Control Engineering Practice, and Guest Associate Editor of Int. Journal of Bifurcations and Chaos, and etc.. His main research interests include theories and applications of complex network and network science, where he has (co-)authored 4 research monographs and more than 150 peer-refereed publications in journals and conferences. Prof. Xiang Li received the IEEE Guillemin-Cauer Best Transactions Paper Award from the IEEE Circuits and Systems Society in 2005, Shanghai Natural Science Award (1st class) in 2008, Shanghai Science and Technology Young Talents Award in 2010, National Science Fund for Distinguished Young Scholar of China in 2014, and other awards and honors.
\end{IEEEbiographynophoto}

\begin{IEEEbiographynophoto}{Yi-Qing Zhang}(S'14)
received the B.S. degrees in electronics engineering from Fudan University, Shanghai, China, in 2007. He is currently a Ph.D. candidate in the Department of Electronic Engineering at Fudan University. His research interests are mostly about dynamics on and of temporal networks, i.e., structure evolution of temporal networks, as well as dynamical processes such as epidemics on temporal networks.
\end{IEEEbiographynophoto}

\begin{IEEEbiographynophoto}{Athanasios V. Vasilakos}
(M'00-SM'11) is currently a Professor with the Kuwait University. He served or is serving as an Editor or/and Guest Editor for many technical journals,such as the IEEE TRANSACTIONS ON NETWORK AND SERVICE MANAGEMENT; IEEE TRANSACTIONS ON CLOUD COMPUTING, IEEE TRANSACTIONS ON INFORMATION FORENSICS AND SECURITY, IEEE TRANSACTIONS ON NANOBIOSCIENCE, IEEE TRANSACTIONS ON CYBERNETICS; IEEE TRANSACTIONS ON INFORMATION TECHNOLOGY IN
BIOMEDICINE;  ACM Transactions on Autonomous and Adaptive Systems; the IEEE JOURNAL ON SELECTED AREAS IN COMMUNICATIONS. He is also General Chair of
the  European Alliances for Innovation(www.eai.eu).
\end{IEEEbiographynophoto}

\end{document}